# Exceptional Points in three-dimensional Nanostructures

Ashok Kodigala, Thomas Lepetit and Boubacar Kanté[*]

*Department of Electrical and Computer Engineering, University of California San Diego, La Jolla, CA 92093-0407, USA*

Exceptional points (EPs) are degeneracies in open wave systems where at least two energy levels and their corresponding eigenstates coalesce. We report evidence of the existence of EPs in 3D plasmonic nanostructures. The systems are composed of coupled plasmonic nanoresonators and can be judiciously and systematically driven to EPs by controlling symmetry-compatible modes via their near-field and far-field interactions. The proposed platform opens the way to the investigation of EPs for enhanced light-matter interactions and applications in communication, sensing and imaging.

Most physical systems are open in nature, *i.e.* energy flows in and out and is exchanged with the environment as radiation and absorption which is in contrast with closed systems where energy stays put and is conserved. Closed systems benefit from the well-established theory for conservative systems, *i.e.* Hermitian systems. A remarkable difference is that in Hermitian systems, eigenmodes do not decay and their corresponding eigenvalues are real whereas in non-hermitian systems eigenmodes do decay and consequently their corresponding eigenvalues are complex [1]. Over the last decade many have sought to bridge the gap between physics of open and closed systems. This renewed attention has underlined one of the fundamental differences between Hermitian and non-Hermitian systems: their singularities. In Hermitian systems, modes couple to induce singularities called diabolical points (DPs), where only the respective eigenvalues are equal whereas for non-Hermitian systems modes couple to induce singularities called exceptional points (EPs), where both eigenvalues and eigenvectors coalesce [2-3].

In conjunction with theoretical inquiries, recent experimental work has given a glimpse of the many promises that an increased understanding of open systems holds. For instance, there has been ample effort in realizing novel photonic devices in the realm of lasers such as: PT-symmetric lasers [4-5], lasers operating near EPs [6], Bound State in Continuum lasers [7-9]. Concurrently, there has also been theoretical progress with strictly passive devices exploiting EPs for a superior sensing scheme that offers enhanced sensitivity [10-11]. Thus far, EPs have been experimentally studied in a variety of physical systems including 2D microwave cavities [12], electronic circuits [13], 2D chaotic optical microcavities [14], and coupled atom-cavity systems [15]. However, to date, exceptional points have not been realized in a fully three-dimensional plasmonic system. This is of importance because it is highly desirable to have a sensitive sub-wavelength sensing system compatible with biologically relevant substances. Plasmons resulting from the interaction between photons and free electrons are ideally suited for biological sensing given the field enhancement and resonance sensitivity to environment.

Here, we report the first evidence of the existence of EPs in an open plasmonic system made of coupled plasmonic nanoresonators. We show that the control of the near-field and far-field interactions lead to a systematic construction of EPs. We subsequently propose a general class of plasmonic architecture exhibiting designer exceptional points.

We consider the plasmonic system based on three coupled nanobars, depicted in Figure 1(a). The dimensions of an individual gold nanobar are chosen such that the fundamental resonance falls in the optical domain at a frequency of 193.5 THz (1.55 µm). Placing these gold nanobars in close proximity couples their individual plasmon modes into hybrid modes as shown in Figure 1(b) [16]. Here, the instantaneous charge profiles of the first three modes are depicted. Intrinsically, the system has reflection symmetry with respect to the xy-plane that bisects the central nanobar and its modes are thus either even or odd. In our case, modes A and C have an even symmetry whereas Mode B has an odd symmetry. Mode A, with eigenfrequency $\omega_A$, has charges in all the bars oscillating in-phase and mode C, with eigenfrequency $\omega_C$, has charges in all bars oscillating out-of-phase. Mode B, $\omega_B$, has no charges in the central bar as seen in Figure 1(b). Therefore, mode A resides at a higher energy (higher frequency) due to all repelling Coulomb interactions and mode C resides at a lower energy (lower frequency) as a result of Coulomb interactions. Lastly, mode B resides between mode A and mode C on the energy scale.

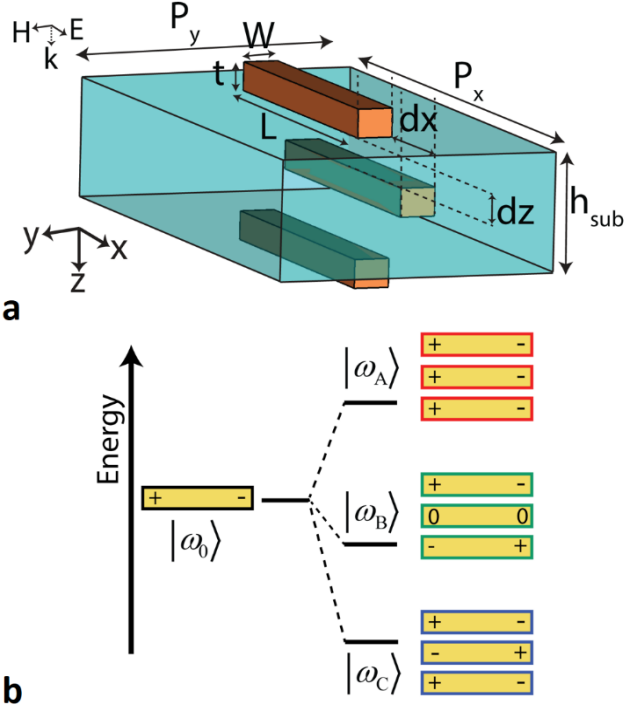

FIG. 1. (a) Physical setup of a unit cell with three paired gold bars, with the middle one separated by a variable distance (dx, dy, dz) with respect to the other two. The dimensions of each nanobar are given by L (450 nm), W (50 nm), and T (40 nm). The periodicity in x and y-directions are given by $P_x$ (800 nm) and $P_y$ (400 nm). The dielectric (SiO2) spacer is shown in blue. The gold bars are described using a Drude model with a plasma frequency ($\omega_p$=1.367x10$^{16}$ rad/sec) and collision frequency ($\omega_c$=6.478x10$^{13}$ rad/sec) [17]. (b) Energy-level diagram describing the plasmon hybridization in the gold-bar system with three modes: $\omega_A$, $\omega_B$, $\omega_C$ where $\omega_A > \omega_B > \omega_C$ for dx=0. $\omega_0$ corresponds to the resonance of an individual bar.

The formation of an EP can be understood as a specific case of mode coupling and can thus be described by Coupled-Mode-Theory (CMT). In this framework, mode coupling is described by a non-Hermitian effective Hamiltonian matrix [18-19].

$$\mathbf{H_{eff}} = \mathbf{H_0} + j\mathbf{\Lambda}_L + j\frac{1}{2}\mathbf{VV}^\dagger \quad (1)$$

Where $\mathbf{H_0}$ is a Hermitian Hamiltonian matrix that describes the system without coupling (closed system). The second term, $j\mathbf{\Lambda}_L$, in the equation represents extraneous losses. In our case, this term accounts for plasmonic losses. The third term, $\mathbf{VV}^\dagger$, describes the coupling with the environment. Hence $\mathbf{H_{eff}}$ describes the full system (open system). Here, the eigenmodes of the system are represented by the complex eigenvalues and eigenvectors of the effective Hamiltonian. Experimentally, however, these eigenvalues are not directly available. Nevertheless, we can measure the scattering spectra and extract eigenvalues as they directly correspond to the complex poles of the scattering spectra [20-21].

An EP is a singularity of the effective Hamiltonian, which arises due to its non-Hermitian nature, at which two modes coalesce [2]. To achieve an EP, both the real and imaginary parts of the eigenvalues (resonance frequency and linewidth) need to coincide simultaneously. For an EP of order 2, such coalescence is dependent on at least two physical parameters [3]. A method is thus needed to select among the geometrical parameters of the system (dx,dy,dz).

For the three-nanobar setup portrayed in Fig. 1, a closed system Hamiltonian can be used for an intuitive understanding of the mode behavior as described below.

$$\mathbf{H_0} = \begin{pmatrix} \omega_0 & \kappa_n & \kappa_{n2} \\ \kappa_n & \omega_0 & \kappa_n \\ \kappa_{n2} & \kappa_n & \omega_0 \end{pmatrix} \quad (2)$$

Here, $\omega_0$ is the uncoupled resonance of an individual nanobar. $\kappa_n$ and $\kappa_{n2}$ are the nearest and next-to-nearest neighbor coupling constants acting between two individual nanobars. We note that this matrix is bisymmetric and hence has eigenvectors that are either symmetric (even) or skew-symmetric (odd) [23]. For a 3x3 $\mathbf{H_0}$, there are always two even (modes A and C) and one odd (mode B) eigenvectors. For the initial three-nanobar setup (dx=0, dy=0, dz=0), $\kappa_n$ is much larger than $\kappa_{n2}$ and the Hamiltonian is almost tridiagonal. This is not favorable for coalescence as even and odd modes are then interlaced. Hence, we need to reduce $\kappa_n$ with respect to $\kappa_{n2}$ to move away from a diagonally dominant Hamiltonian (1$^{st}$ constraint). Besides, since even and odd modes do not couple, we are only interested in the coalescence of the two even modes. Therefore, we seek a parameter that does not introduce coupling between even and odd modes, *i.e.* does not break the system's mirror symmetry (2$^{nd}$ constraint). Both constraints can be met by shifting the middle bar along the x-direction [21-22].

Since plasmonic losses in these identical nanobars are represented by a scalar matrix, the losses only contribute an overall complex shift. Moreover, the coupling to the environment adds to the imaginary part of the eigenvalues.

$$\lambda_i = \omega_i + j\gamma_L^i - j\frac{1}{2}\frac{\boldsymbol{x}_L^i \mathbf{VV}^\dagger \boldsymbol{x}_R^i}{\boldsymbol{x}_L^i(\mathbf{H_0} + j\mathbf{\Lambda}_L)\boldsymbol{x}_R^i} \quad i \in [\![a,b,c]\!] \quad (3)$$

Here, $\boldsymbol{x}_L$ and $\boldsymbol{x}_R$ are the left and right eigenvectors respectively. For a sufficient shift, dx and dz, mode A and mode C become degenerate (complex eigenvalue).

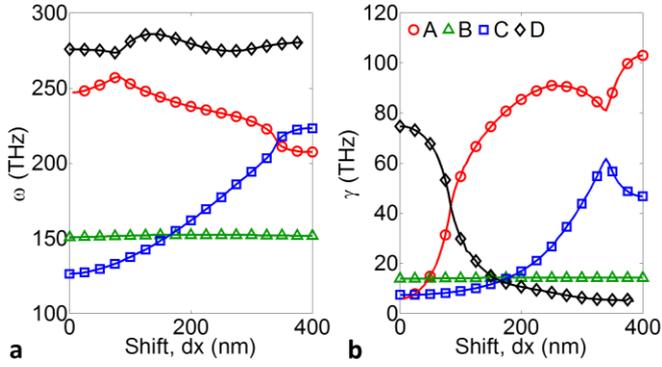

FIG. 2. Resonance information in the form of complex poles extracted from scattering parameters and plotted as a function of shift 'dx' (middle-bar) for $p_x$=800 nm and $dz$=60 nm. (a) Resonance frequency of modes A (○), B (△), C (□) and higher order mode D (◊) with varying 'dx' and their corresponding (b) linewidths. There is observable coupling between neighboring modes that share a symmetry, i.e. mode A with C at dx=350 nm and mode A with D at dx=80 nm. Mode B is unperturbed by both the shift and neighboring modes due to its symmetry. Coupling of modes A and C is of interest for this parameter set as the resonance frequency cross with 'dx' and linewidths experience an avoided resonance crossing.

We now numerically examine the effect of shifting the middle bar in the x-direction on all three modes of the coupled plasmonic system (see Figure 2). As the middle bar is progressively displaced, the repelling forces associated with mode A weaken to become attractive. Similarly, the attractive forces of mode C weaken to become repulsive. Lastly, the Coulomb forces associated with mode B remain constant with shift of the central bar as there is no field present in this bar. This behavior is noticeable in the resonances of this system as seen in Figure 2(a). Mode A moves to lower frequencies with shift and mode C moves to higher frequencies with shift whereas mode B remains unperturbed. Due to the presence of a higher-order resonance (mode D), also with an even symmetry, mode A does not monotonously decrease with shift. For values of 'dx' below 80 nm, mode A increases in frequency with shift due to coupling to mode D. As evident from the coupling between even modes A and D around dx=100 nm and between modes A and C at dx=340 nm, neighboring resonances of shared symmetry couple to each other. Having an odd symmetry, mode B never couples to any of the even modes. The coupling between modes is further evident in their linewidth behavior as seen in Figure 2b. As modes A and D are avoided in frequency at dx=80 nm, their respective linewidths cross. Similarly, modes A and C cross in frequency at dx=340 nm and their linewidths exhibit an avoided resonance crossing. In terms of the near-field coupling terms, at no shift, i.e. dx=0, $\kappa_n$ is the dominant coupling term. With an increase in dx, $\kappa_n$ weakens with respect to $\kappa_{n2}$. It is precisely this interplay that forces the eigenvalues associated with modes A and C to converge towards one another, which is mandatory for engineering an EP. Note that the present system is not exactly at an EP.

In the close vicinity of an order-2 EP, the effective Hamiltonian of this system can be written in its reduced form as a 2x2 matrix considering only the two concerned even modes [2].

$$\mathbf{H}_{eff} = \begin{bmatrix} \omega_A & 0 \\ 0 & \omega_C \end{bmatrix} + j \begin{bmatrix} \gamma_A & \sqrt{\gamma_A \gamma_C} \\ \sqrt{\gamma_A \gamma_C} & \gamma_C \end{bmatrix} \quad (4)$$

As stated earlier, realization of an EP via two modes requires at least two physical parameters. The two parameters used for the above system to reach an EP are a shift, dx, in the central bar and the inter-spacing between nanobars, dz, in the z-direction where both parameters influence $\kappa_n$ and $\kappa_{n2}$. By performing detailed full-wave finite element simulations, we present here a numerical proof of an EP in our nanobars system (see Figure 3). An EP occurs at a frequency of ~212 THz for a 345 nm lateral shift of the middle bar and an inter-particle spacing close to 61 nm. For dz=61 nm, the two resonance frequencies ($\omega_A$, $\omega_C$) cross each other with increasing shift, dx, and the linewidths ($\gamma_A$, $\gamma_C$) avoid each other as seen in Fig. 3(a). Conversely, for dz=61.5 nm, the linewidths cross and frequencies are avoided as seen in Fig. 3(b). For a value between 61 and 61.5 nm, there is a definite occurrence of an EP singularity where both resonance frequencies and linewidths coalesce.

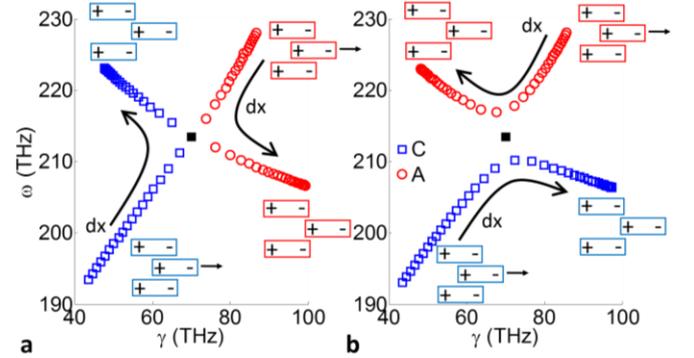

FIG. 3. Resonances approaching an exceptional point (black ■) plotted in the complex plane (γ+jω) for modes A (○) and C (□) as a function of 'dx' (300 to 400 nm) for two different values of inter-bar spacing, dz=61 and 61.5 nm (increasing 'dx' indicated by arrows). (a) For dz=61 nm, the resonance frequencies of modes A and C cross as the center bar is shifted (dx) but the linewidths are avoided whereas (b) for dz=61.5 nm, the linewidths cross and the resonance frequencies are avoided. An EP singularity occurs at a value of 'dz' between 61 and 61.5 nm for a dx of ~345 nm where both resonance frequencies and linewidths coalesce.

Another indication of an occurrence of an EP lies with the complex residues of the corresponding complex poles associated with the resonances [24, 25]. In the case of the three-nanobar system, both the real and imaginary components of the residues diverge as one approaches the EP (see Figure 4(a),(b)). As the EP is approached from the left, or increasing dx, the real parts diverge and similarly the imaginary parts diverge as the EP is approached from the right. However, the sum of the residues for both the real and imaginary remain finite (see Figure 4(c), (d)) [26].

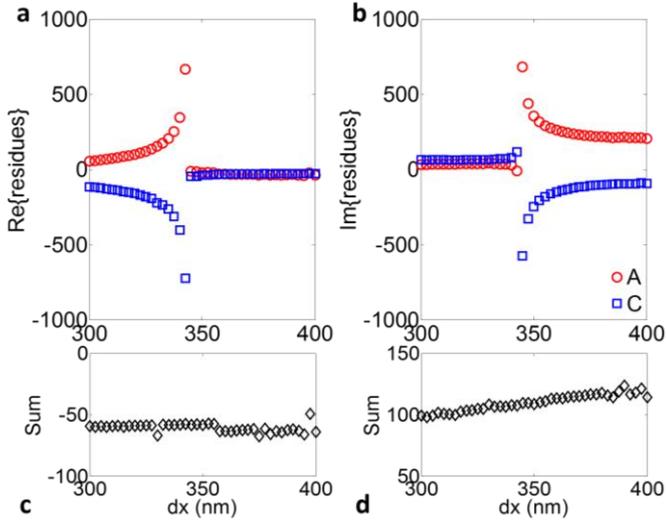

FIG. 4. Residues of the corresponding modes A (○) and C (□) as a function of shift, dx, for dz=61 nm. (a) Real and (b) Imaginary parts of the residues diverging when approaching the EP (dx=345 nm). Sum of the (c) real and (d) imaginary parts of the residues which remain finite.

Furthermore, an EP is not exclusive to the three-bar system. An EP can also be realized in systems with more plasmonic resonators in a given unit cell. Here, we address the general case of having an odd number of bars (N=2n+1) in a unit cell and once again guided by an NxN closed system Hamiltonian. In general, for such a matrix of order N, there are $\lceil N/2 \rceil$ even and $\lfloor N/2 \rfloor$ odd eigenvectors. These eigenvectors are alternately even and odd with eigenvalues arranged in descending order given that the eigenvalues are distinct. The resulting eigenvectors of eigenvalues (see Eq. 3) can be expressed as

$$(\mathbf{u} \quad \alpha \quad +\mathbf{Ju})^T \quad \text{(even eigenvectors)} \quad (5)$$
$$(\mathbf{u} \quad 0 \quad -\mathbf{Ju})^T \quad \text{(odd eigenvectors)}$$

Here, J is the exchange matrix [23]. Note for an odd eigenvector, there is no excitation or field in the central bar as was the case for Mode B earlier.

As an example, we take the case with five coupled bars (n=2) described by 5x5 Hamiltonian, $\mathbf{H_0}$, written as follows when all bars are perfectly aligned in the z-direction, i.e. dx=0.

$$\mathbf{H_0} = \begin{pmatrix} \omega_0 & \kappa_n & \kappa_{n2} & 0 & 0 \\ \kappa_n & \omega_0 & \kappa_n & \kappa_{n2} & 0 \\ \kappa_{n2} & \kappa_n & \omega_0 & \kappa_n & \kappa_{n2} \\ 0 & \kappa_{n2} & \kappa_n & \omega_0 & \kappa_n \\ 0 & 0 & \kappa_{n2} & \kappa_n & \omega_0 \end{pmatrix} \quad (6)$$

Here, we can neglect the coupling terms $\kappa_{n3}$ and $\kappa_{n4}$ as they are simply dominated by $\kappa_n$ and $\kappa_{n2}$. Similar to the three-bar case, we must choose physical parameters to modify so as to weaken $\kappa_n$ and strengthen $\kappa_{n2}$. In order to retain the bisymmetric nature of the Hamiltonian, we note that all nearest-neighbor and next-to-nearest-neighbor coupling terms need to be the same as you modify the geometry of the system in accordance with the two constrains outlined earlier. Therefore, we concurrently shift the top, middle and the bottom bars in the x-direction which satisfies this condition and appropriately modifies $\kappa_n$ and $\kappa_{n2}$. For an order N=5, there are three even and two odd eigenvectors. For an EP, we focus our attention on interaction between two of the even modes. The two parameters are still the inter-spacing, dz, along the z-direction and shift, dx (see Figure 5). Similar to the three-bar case, we observe resonances crossing in frequency and an avoided crossing in linewidths as evidence of an EP. An EP occurs at a frequency of ~227 THz for a 345 nm lateral shift of the bars and an inter-particle spacing, dz, close to 42 nm. This approach is general and can be utilized to engineer an EP in coupled nanoresonator structures which can be physically realized [27].

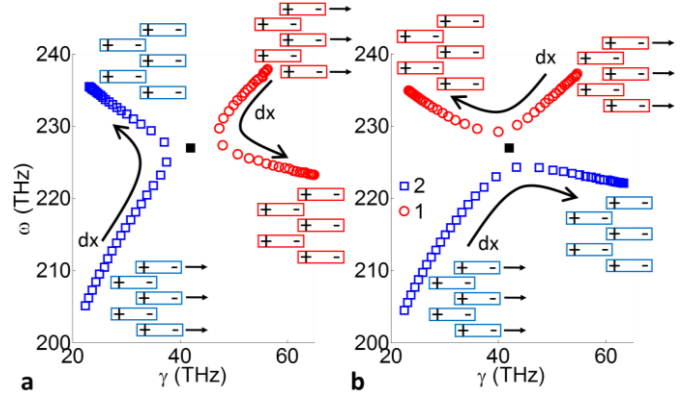

FIG. 5. Realization of an exceptional point in a system with 5 bars (2n+1 with n=2) with top, middle and bottom bars shifted by dx (300 to 400 nm) for dz=42 nm and 43 nm. Mode 1 (○) and Mode 2 (□): two modes of shared symmetry interact to form an EP (■) at a value of d between 42 and 43 nm for a dx of ~345 nm. (a) For dz=42 nm, the resonance frequencies of modes 1 and 2 cross as the bars, indicated by arrows, are shifted (dx) but the linewidths are avoided whereas (b) for dz=43 nm, the linewidths cross and the resonance frequencies are avoided.

We have demonstrated the existence of exceptional points in three dimensional systems of coupled plasmonic nanostructures. The EP is constructed by coalescing symmetry-compatible modes and its existence is further evident from the diverging complex residues in the vicinity of the EP singularity. A thorough discussion on the importance of mode symmetries for EPs was presented.

The general approach to designing EPs in systems of coupled resonators proposed here can be used to construct EPs of higher order in physical systems where more than two modes coalesce. These ideas could be applied to other areas of wave physics such as acoustic and matter waves. We believe this work paves the way to the experimental observation of exceptional points in various physical systems and will foster further research towards unprecedented sensing schemes.

*bkante@ucsd.edu

1. N. Moiseyev, Non-Hermitian Quantum Mechanics (Cambridge University Press, United Kingdom 2011).